\documentclass[12pt]{article}
\usepackage{latexsym}
\usepackage{amssymb}
\usepackage{amsmath}
\usepackage{mathrsfs}
\usepackage{url}
\usepackage{cite}

\newtheorem{theorem}{Theorem}[section]
\title{\Large \textbf{Recursion operators and bi-Hamiltonian structure of the general heavenly equation}}
\author{\large\textbf{M. B. Sheftel$^1$, A. A. Malykh$^2$ and D. Yaz{\i}c{\i}$^3$}\\[5mm]
 \normalsize $^1$~Department of Physics, Bo\u{g}azi\c{c}i University\\ \normalsize 34342 Bebek, Istanbul, Turkey\\
\normalsize $^2$~Department of Numerical Modelling,\\ \normalsize Russian State Hydrometeorlogical University,
\\ \normalsize Malookhtinsky pr. 98, 195196 St. Petersburg, Russia\\ \normalsize
$^3$~Department of Physics, Y{\i}ld{\i}z Technical University\\ \normalsize 34220 Esenler, Istanbul, Turkey\\ \normalsize
E-mail: mikhail.sheftel@boun.edu.tr, andrei-malykh@mail.ru \\ \normalsize and yazici@yildiz.edu.tr}
\date{}

\begin{document}
\maketitle

\begin{abstract}
 We discover two additional Lax pairs and three nonlocal recursion operators for symmetries of the general heavenly equation introduced by Doubrov and Ferapontov. Converting the equation to a two-component form, we obtain Lagrangian and Hamiltonian structures of the two-component general heavenly system.
We study all point symmetries of the two-component system and, using the inverse Noether theorem in the Hamiltonian form, obtain all the integrals of motion corresponding to each variational (Noether) symmetry. We discover that in the two-component form we have only a single nonlocal recursion operator. Composing the recursion operator with the first Hamiltonian operator we obtain second Hamiltonian operator. We check the Jacobi identities for the second Hamiltonian operator and compatibility of the two Hamiltonian structures using P. Olver's theory of functional multi-vectors. Our well-founded conjecture is that P. Olver's method works fine for nonlocal operators. We show that the general heavenly equation in the two-component form is a bi-Hamiltonian system integrable in the sense of Magri. We demonstrate how to obtain nonlocal Hamiltonian flows generated by local Hamiltonians by using formal adjoint recursion operator.
\end{abstract}



\section{Introduction}
\label{sec-intro}

In paper \cite{fer}, Doubrov and Ferapontov introduced the general heavenly equation (GHE)
\begin{equation}
  \alpha u_{12}u_{34} + \beta u_{13}u_{24} + \gamma u_{14}u_{23} = 0,\qquad \alpha + \beta + \gamma = 0
  \label{gener}
\end{equation}
where $\alpha, \beta$ and $\gamma$ are arbitrary constants satisfying one linear relation given above. Here $u=u(z_1,z_2,z_3,z_4)$
is a holomorphic function of four complex variables. They also obtained the Lax pair for equation (\ref{gener})
\begin{eqnarray}
   X_1 &=& u_{34}\partial_1 - u_{13}\partial_4 + \gamma\lambda(u_{34}\partial_1 - u_{14}\partial_3),
\nonumber   \\
   X_2 &=& u_{23}\partial_4 - u_{34}\partial_2 + \beta\lambda(u_{34}\partial_2 - u_{24}\partial_3),
  \label{lax}
\end{eqnarray}
where $\partial_i$ means $\partial/\partial z_i$, while $u_{ij}=\partial^2 u/\partial z_i\partial z_j$.

There are only few examples of multi-dimensional integrable systems. The so-called heavenly equations make up an important class of such integrable systems since they are obtained by a reduction of the Einstein equations with Euclidean (and neutral) signature for (anti-)self-dual gravity which includes the theory of gravitational instantons. All of these equations are of Monge-Amp\`ere type, so that the only nonlinear terms are Hessian $2\times 2$ determinants.
General heavenly equation is an important example of such equations. An explicit description of ASD Ricci-flat vacuum metric governed by GHE, null tetrad and basis 1-forms for this metric were obtained in our paper \cite{genheav}. Recently, Bogdanov \cite{bogdan} showed important relations between GHE and heavenly equations of Pleba\'nski and developed $\bar{\partial}$-dressing scheme for GHE in the context of the inverse scattering method. This stresses the importance of further study  of the Doubrov-Ferapontov's general heavenly equation.

In this paper, we obtain recursion operators for symmetries of GHE (\ref{gener}) and discover its bi-Hamiltonian structure in a two-component form, where the single second-order PDE (\ref{gener}) is presented as an evolutionary system of two first-order PDEs with two unknowns. Therefore, by the theorem of Magri \cite{magri} this general heavenly (GH) system is completely integrable.

While completing this paper, we became aware of the preprint \cite{Artur} by A. Sergyeyev where he discovers a recursion operator for the one-component version of GHE, which coincides with the first one of our recursion operators, as an example of application of his general method for constructing recursion operators for dispersionless integrable systems.

If $\varphi$ is a symmetry characteristic for (\ref{gener}), it satisfies the symmetry condition, ``linearization'' of equation (\ref{gener})
\begin{equation}
\hat A \varphi = \alpha(u_{34}\varphi_{12} + u_{12}\varphi_{34}) + \beta(u_{24}\varphi_{13} + u_{13}\varphi_{24})
+ \gamma(u_{23}\varphi_{14} + u_{14}\varphi_{23}) = 0.
\label{symcond}
\end{equation}
A recursion operator maps any symmetry again into a symmetry and, as a consequence, it commutes with the operator $\hat A$ on solutions. For all other heavenly equations in the classification \cite{fer} of Doubrov and Ferapontov the symmetry condition has a two-dimensional divergence form which allows us to introduce partner symmetries \cite{shma,mns,MNS,manush,shya}, a powerful tool for finding recursion operators and noninvariant solutions \cite{mash,ShMa} which are necessary for the construction of the famous gravitational instanton $K3$. It is easy to check that the symmetry condition (\ref{symcond}) for the general heavenly equation cannot be presented in a two-dimensional divergence form but it can be presented in a three-dimensional divergence form.

Therefore, the method of partner symmetries does not work any more for GHE, so that we have to use here a different approach in order to find a recursion operator which could be regarded as a generalization of the method of partner symmetries. This approach is based on splitting each of the two Lax operators with respect to the spectral parameter $\lambda$ in two operators and multiplying the first operator by the inverse of the second operator in each pair, obtained by the splitting in $\lambda$. This method was presented in \cite{Artur} using somewhat more geometric language. The idea of obtaining recursion operators from Lax pairs was used in 2003 in our paper on partner symmetries of the complex Monge-Amp\`ere equation \cite{mns}.

For a single-component equation (\ref{gener}), we discover three Lax pairs, related by discrete symmetries of both GHE and its symmetry condition, and three recursion operators corresponding to them. However, a two component form of the equation (\ref{gener}) is not invariant under these permutations of indices, so two other recursion operators are related to two other 2-component systems. We do not consider them here because they are obtained from our two-component system just by the permutations of indices.

Another important property of heavenly equations is that they can be presented in a two-component evolutionary form as bi-Hamiltonian systems \cite{nns,nsky,sy,Ya}. We show here that GHE also possesses this property. In a two-component form we construct a Lagrangian for this system and discover its symplectic and Hamiltonian structure. We obtain all point symmetries of this system and, using its Hamiltonian structure, apply the inverse Noether theorem to derive Hamiltonians generating all symmetry flows. Hamiltonians of the symmetry flows which commute with the GH flow are integrals of the motion for the GH system. Each such Hamiltonian is also conserved by all the symmetry flows that commute with the symmetry generated by this Hamiltonian. This procedure works only for variational (Noether) symmetries.
Composing the recursion operator in a $2\times 2$ matrix form with
the Hamiltonian operator $J_0$ we generate a candidate for the second Hamiltonian operator $J_1 = RJ_0$. The property of $J_1$ to be Hamiltonian operator is equivalent to the recursion operator being hereditary (Nijenhuis) \cite{ff,sheftel}. Since this property of our $R$ is not known, we check directly the Jacobi identities for $J_1$, which is obviously skew-symmetric, and the compatibility of the two Hamiltonian operators $J_0$ and $J_1$.
We find the corresponding Hamiltonian density $H_0$ such that the original general heavenly flow is generated by the action of $J_1$ on variational derivatives of the Hamiltonian functional $\cal{H}$$_0$, so that the GH system turns out to be a bi-Hamiltonian system. We demonstrate how applying the formal adjoint recursion operator $R^\dagger$ we can generate higher flows which are nonlocal symmetries of the system. We show how further local Hamiltonians can be constructed which generate nonlocal Hamiltonian flows.

In section \ref{sec-recurs}, we obtain two more Lax pairs for GHE in addition to the Doubrov-Ferapontov Lax pair. We show how to use these three Lax pairs for constructing three nonlocal recursion operators for GHE.

In section \ref{sec-twocomp}, we present the general heavenly equation in a two-component evolutionary form and obtain a Lagrangian for this GH system.

In section \ref{sec-Ham1}, we discover symplectic and Hamiltonian structure of the GH system.

In section \ref{sec-symint}, we obtain all point Lie symmetries of the GH system and show how the corresponding Hamiltonians of the variational symmetry flows can be obtained from the inverse Noether theorem in a Hamiltonian form. These Hamiltonians are integrals of the motion along the original GH flow for all the variational symmetry flows that commute with the GH flow. Each Hamiltonian is also conserved along all the symmetry flows that commute with the flow generated by this Hamiltonian.

In section \ref{sec-recurs_twocomp}, we derive a nonlocal recursion operator for the two component GH system.

In section \ref{sec-2ndH}, composing the recursion operator with the first Hamiltonian operator $J_0$ we obtain the second nonlocal Hamiltonian operator $J_1$. We also find the corresponding Hamiltonian density which generates the GH system with the aid of the second Hamiltonian operator. Therefore, we obtain a bi-Hamiltonian representation for the general heavenly equation in a two-component form.

In section \ref{sec-jacobi}, we prove the Jacobi identities for the second Hamiltonian operator $J_1$ and compatibility of the two Hamiltonian structures
$J_0$ and $J_1$ using the theory of functional multivectors by P. Olver \cite{olv}. So far the nonlocal Hamiltonian operators for equations of the Monge-Amp\`ere type involved only operators inverse to total derivatives. Now, the operator $J_1$ is \textit{essentially nonlocal}, i.e. it involves the operator inverse to the linear combination of total derivatives with coefficients depending on the derivatives of unknown $u$. We show how P. Olver's theory works nicely even for this more complicated case. The applicability of P. Olver's method to nonlocal Hamiltonian operators seems to be a well-founded conjecture though a rigorous generalization is still awaited.

In section \ref{sec-higher}, we demonstrate how to use the formal adjoint recursion operator $R^\dagger$ to generate an infinite number of higher nonlocal symmetry flows and obtain their Hamiltonians which are integrals of the motion.

\section{Recursion operators}
\setcounter{equation}{0}
 \label{sec-recurs}

We introduce the following first-order differential operators from which the Lax operators (\ref{lax}) are constructed
\begin{eqnarray}
L_{14(3)} = u_{34}D_1 - u_{13}D_4, \quad  L_{13(4)} = u_{34}D_1 - u_{14}D_3,\nonumber
 \\  L_{24(3)} = u_{34}D_2 - u_{23}D_4,\quad L_{23(4)} = u_{34}D_2 - u_{24}D_3
 \label{Lops}
\end{eqnarray}
where $D_i$ denotes total derivative with respect to $z_i$. We could restrict ourselves just by the general definition
$L_{ij(k)} = {u_{jk}D_i - u_{ik}D_j}$, so that $L_{ji(k)} = - L_{ij(k)}$ but explicit expressions for different values of $i,j,k$ are given for reader's convenience. Lax operators take the form
\begin{equation}
 X_1 = L_{14(3)} + \lambda\gamma L_{13(4)}, \qquad X_2 = - L_{24(3)} + \lambda\beta L_{23(4)}.
\label{LaxComp}
\end{equation}
The symmetry condition (\ref{symcond}), where we use $\alpha=-(\beta+\gamma)$
\begin{eqnarray}
\hat A \varphi = \{\beta(L_{14(2)}D_3 - L_{14(3)}D_2) + \gamma(L_{24(1)}D_3 - L_{24(3)}D_1)\}\varphi = 0 \nonumber \\
\equiv \{\beta(D_3L_{14(2)} - D_2L_{14(3)}) + \gamma(D_3L_{24(1)} - D_1L_{24(3)})\}\varphi = 0
\label{symcon1}
\end{eqnarray}
contains two more operators
\begin{equation}
 L_{14(2)} = u_{24}D_1 - u_{12}D_4,\quad L_{24(1)} = u_{14}D_2 - u_{12}D_4 .
\label{Ladd}
\end{equation}

To arrive at two different Lax pairs, we apply two permutations of indices which leave invariant both the equation (\ref{gener}) and its
symmetry condition (\ref{symcond}) but which do change the Lax pair and recursion operators.
The permutation $1\leftrightarrow3,\; 2\leftrightarrow4$ yields the second Lax pair
\begin{equation}
 X_1^{(2)} = L_{23(1)} + \lambda\gamma L_{13(2)},\qquad X_2^{(2)} = L_{24(1)} - \lambda\beta L_{14(2)}
\label{lax2}
\end{equation}
where
\begin{equation}
  L_{13(2)} = u_{23}D_1 - u_{12}D_3,\qquad L_{23(1)} = u_{13}D_2 - u_{12}D_3
\label{opers2}
\end{equation}
and operators $L_{14(2)},\; L_{24(1)}$ are defined in (\ref{Ladd}). In (\ref{lax2}) we have skipped the overall minus in $X_1^{(2)}$.

The permutation $1\leftrightarrow4,\; 2\leftrightarrow3$ yields the third Lax pair
\begin{equation}
 X_1^{(3)} = L_{41(2)} + \lambda\gamma L_{42(1)},\qquad X_2^{(3)} = - L_{31(2)} + \lambda\beta L_{32(1)}
\label{lax3}
\end{equation}
where
\begin{eqnarray}
L_{41(2)} = u_{12}D_4 - u_{24}D_1,\qquad L_{42(1)} = u_{12}D_4 - u_{14}D_2,\nonumber \\
L_{31(2)} = u_{12}D_3 - u_{23}D_1,\qquad L_{32(1)} = u_{12}D_3 - u_{13}D_2 .
\label{opers3}
\end{eqnarray}

We note that the symmetry condition (\ref{symcon1}) has a three-dimensional divergence form and therefore our definition of
partner symmetries, which requires a two-dimensional divergence form of the symmetry condition \cite{shma}, does not work here,
so that we need here a different approach to obtain a recursion operator which is applied to the proof of the following theorem.
We note that recursion operators for partner symmetries of the heavenly equations, which we studied before, are composed from the operators obtained by splitting the Lax operators with respect to spectral parameter $\lambda$ and multiplying the first operator by the inverse of the second operator in each pair, that was obtained by splitting in $\lambda$ \cite{mns,MNS,manush,nsky,sy}.
\begin{theorem}
The general heavenly equation admits the following three different Lax pairs (\ref{lax_1}), (\ref{lax_2}), (\ref{lax_3}) and three respective
recursion operators defined by the relations (\ref{ro1}), (\ref{ro2}), (\ref{ro3})
\begin{subequations}
\begin{equation}
 X_1^{(1)} = L_{14(3)} + \lambda\gamma L_{13(4)}, \quad X_2^{(1)} = - L_{24(3)} + \lambda\beta L_{23(4)}
\label{lax_1}
\end{equation}
\begin{equation}
  L_{14(3)}\varphi = \gamma L_{13(4)}\psi,\qquad L_{24(3)}\varphi = - \beta L_{23(4)}\psi.
\label{ro1}
\end{equation}
\end{subequations}
\begin{subequations}
\begin{equation}
 X_1^{(2)} = L_{23(1)} + \lambda\gamma L_{13(2)},\quad X_2^{(2)} = L_{24(1)} - \lambda\beta L_{14(2)}
\label{lax_2}
\end{equation}
\begin{equation}
  L_{23(1)}\varphi = \gamma L_{13(2)}\psi,\qquad L_{24(1)}\varphi = - \beta L_{14(2)}\psi.
\label{ro2}
\end{equation}
\end{subequations}
\begin{subequations}
\begin{equation}
 X_1^{(3)} = L_{41(2)} + \lambda\gamma L_{42(1)},\quad X_2^{(3)} = - L_{31(2)} + \lambda\beta L_{32(1)}
\label{lax_3}
\end{equation}
\begin{equation}
  L_{41(2)}\varphi = \gamma L_{42(1)}\psi,\qquad L_{31(2)}\varphi = - \beta L_{32(1)}\psi.
\label{ro3}
\end{equation}
\end{subequations}
\end{theorem}

\underline{\textbf{Proof}}\\
The Lax pair (\ref{lax_1}) is known from the paper \cite{fer} while two other Lax pairs are obviously true since they are obtained from (\ref{lax_1}) by the permutations of indices which do not change the equation (\ref{gener}).

To prove that the relations (\ref{ro1}) indeed represent a recursion operator, we analyze their integrability conditions.
We have two sets of integrability conditions for (\ref{ro1})
\begin{equation}
  [L_{24(3)},L_{14(3)}]\varphi = (\gamma L_{24(3)}L_{13(4)} + \beta L_{14(3)}L_{23(4)})\psi
\label{int1}
\end{equation}
\begin{equation}
 \beta\gamma [L_{23(4)},L_{13(4)}]\psi = (\beta L_{23(4)}L_{14(3)} + \gamma L_{13(4)}L_{24(3)})\varphi .
\label{int2}
\end{equation}
In condition (\ref{int1}), the commutator in the left-hand-side expands as
\[[L_{24(3)},L_{14(3)}]\varphi = \frac{1}{u_{34}}\{(u_{34}u_{234} + u_{23}u_{344})L_{14(3)} + (u_{34}u_{134} - u_{13}u_{344})L_{24(3)}\}\varphi\]
which, with the use of equations (\ref{ro1}) converts to derivatives of $\psi$. In condition (\ref{int2}), the commutator in the left-hand-side expands as
\[[L_{23(4)},L_{13(4)}]\psi = \frac{1}{u_{34}}\{(u_{34}u_{234} - u_{24}u_{344})L_{13(4)} - (u_{34}u_{134} - u_{14}u_{334})L_{23(4)}\}\psi\]
which again with the use of (\ref{ro1}) converts to derivatives of $\varphi$. Thus, the condition (\ref{int1}) becomes the equation for $\psi$ only,
which can be straightforwardly checked to coincide with
$\hat A\psi = 0$, while the condition (\ref{int2}) is the equation $\hat A\varphi = 0$ for $\varphi$ only. Therefore, the integrability conditions of equations
(\ref{ro1}) are symmetry conditions for $\psi$ and $\varphi$, which means that both $\psi$ and $\varphi$ are symmetry characteristics for GHE (\ref{gener}). Hence, the relations (\ref{ro1}) are recursion relations between symmetries $\psi$ and $\varphi$ and, consequently, the relations (\ref{ro1}) determine a recursion operator. Since two other relations (\ref{ro2}) and (\ref{ro3}) are obtained from the recursion relations (\ref{ro1}) by permutations of indices, which
leave invariant the equation (\ref{gener}) and its symmetry condition, these two relations also obviously determine recursion operators. $\mathbf{\Box}$

We note that the recursion relations (\ref{ro1}) look like an auto-B\"acklund transformation between the symmetry conditions $\hat A\varphi = 0$ and $\hat A\psi = 0$. Later, we have found that the same observation was made earlier in \cite{Artur} in a more general context. We have checked that our recursion relations (\ref{ro1}) coincide with the ones obtained somewhat earlier by A. Sergyeyev in the version 1 of \cite{Artur}.
The approach we used here can be regarded as a generalization of the concept of partner symmetries.

\section{Two-component evolutionary form of the general heavenly\\ equation}
\setcounter{equation}{0}
 \label{sec-twocomp}

In order to discover Hamiltonian structure of GHE we need to convert it to a two-component evolutionary form.
For this purpose, we transform $z_1, z_2$ into the ``time'' and ``space'' variables $t$ and $x$
\[t = z_1 + z_2,\quad x = z_1 - z_2,\quad y = z_3,\quad z = z_4.\]
  GHE (\ref{gener}) becomes
\begin{equation}
 \alpha(u_{tt} - u_{xx})u_{yz} + \beta(u_{ty} + u_{xy})(u_{tz} - u_{xz}) + \gamma(u_{ty} - u_{xy})(u_{tz} + u_{xz}) = 0.
\label{transgen}
\end{equation}
To convert (\ref{transgen}) into an evolutionary system, we define the second component as $v = u_t$ with the following final result for (\ref{transgen})
\begin{equation}
 \left\{\begin{array}{lcl}
 u_t = v \\ \displaystyle
 v_t = \frac{1}{u_{yz}}\left[u_{xx}u_{yz} - u_{xy}u_{xz} + v_yv_z + b(v_yu_{xz} - v_zu_{xy})\right]
 \end{array} \right.
\label{twocomp}
\end{equation}
where $\displaystyle b=\frac{\beta-\gamma}{\alpha}$ is the single parameter of the general heavenly flow (\ref{twocomp}) constructed from the coefficients of GHE (\ref{gener}).
It is easy to check that (\ref{twocomp}) are Euler-Lagrange equations with the Lagrangian density
\begin{equation}
  L = \left(vu_t - \frac{v^2}{2}\right)u_{yz} - \frac{1}{2}u_yu_zu_{xx} + \frac{b}{3} u_t(u_zu_{xy} - u_yu_{xz}).
\label{Lagrange}
\end{equation}

\section{First Hamiltonian structure}
\setcounter{equation}{0}
 \label{sec-Ham1}

We define momenta
\begin{equation}
 \Pi_u = \frac{\partial L}{\partial u_t} = vu_{yz} + b(u_zu_{xy} - u_yu_{xz}), \quad  \Pi_v = \frac{\partial L}{\partial v_t} = 0
\label{moment}
\end{equation}
which shows that the Lagrangian (\ref{Lagrange}) is degenerate since the momenta cannot be inverted for the velocities. Therefore, following Dirac \cite{dirac}
we impose (\ref{moment}) as constraints
\begin{equation}
  \Phi_u = \Pi_u - vu_{yz} - b(u_zu_{xy} - u_yu_{xz}), \quad \Phi_v = \Pi_v
\label{constr}
\end{equation}
and calculate the Poisson bracket of the constraints \newline $K_{ij} = [\Phi_i(x,y,z),\Phi_j(x',y',z')]$, where $i,j = 1,2$, $u_1=u, u_2=v$,\\
$\Phi_1 = \Phi_u,\; \Phi_2 = \Phi_v$ using
\[[\Pi_i(x,y,z),u_k(x',y',z')] = \delta_i^k \delta(x - x')\delta(y-y')\delta(z - z').\]
The result in a matrix form reads
\begin{equation}
  K = \left(
  \begin{array}{cc}
    b(u_{xz}D_y - u_{xy}D_z) + v_zD_y + v_yD_z + v_{yz}, & - u_{yz} \\
      u_{yz}                                             &       0
  \end{array}
  \right)
\label{Kmatr}
\end{equation}
which is obviously a skew symmetric matrix. The symplectic structure is defined in terms of (\ref{Kmatr}) by the differential two-form
$\omega = \frac{1}{2}du_i\wedge K_{ij}du_j$ with the final result
\begin{eqnarray}
  \omega &=& \frac{b}{2}(u_{xz}du\wedge du_y - u_{xy}du\wedge du_z) + \frac{1}{2}(v_zdu\wedge du_y + v_ydu\wedge du_z)\nonumber
   \\ & &\mbox{} - u_{yz}du\wedge dv.
\label{omega}
\end{eqnarray}
It is easy to check that the two-form (\ref{omega}) is closed, $d\omega=0$, up to a total divergence and hence it determines a symplectic structure. The inverse to the symplectic operator is the Hamiltonian operator $J_0$ because the closed condition for $\omega$ is equivalent to satisfaction of Jacobi identities for $J_0$ \cite{olv}. Thus, we obtain the first Hamiltonian operator in the form
\begin{equation}
  J_0 = K^{-1} = \frac{1}{\sqrt{\det K_{ij}}}\left(
  \begin{array}{cc}
         0 & u_{yz} \\
  - u_{yz} & K_{11}
  \end{array}
  \right)\frac{1}{\sqrt{\det K_{ij}}}
\label{h1}
\end{equation}
where $\det K_{ij} = u_{yz}^2$. In a final form, the first Hamiltonian operator reads
\begin{equation}
  J_0 = \left(
  \begin{array}{cc}
  0     &  \displaystyle\frac{1}{u_{yz}} \\
 \displaystyle - \frac{1}{u_{yz}} & J_0^{22}
  \end{array}
  \right)
\label{Ham1}
\end{equation}
where
\begin{equation}
  J_0^{22} = \frac{1}{u_{yz}}\left[b(u_{xz}D_y - u_{xy}D_z) + v_zD_y + v_yD_z + v_{yz}\right]\frac{1}{u_{yz}},\qquad b=\frac{\beta-\gamma}{\alpha}.
\label{J022a}
\end{equation}
We note that $J_0$ is obviously skew symmetric.

The Hamiltonian density for the system (\ref{twocomp}) corresponding to the Lagrangian (\ref{Lagrange}) is determined by
\begin{equation}
   H_1 = \Pi_u u_t + \Pi_v v_t - L = \frac{1}{2}\left(v^2u_{yz} + u_yu_zu_{xx}\right)\; \iff\; H_1 = \frac{1}{2}\left(v^2 + u_x^2\right)u_{yz}
\label{H1}
\end{equation}
the two expressions being equivalent because their difference is a total divergence.
The system (\ref{twocomp}) in the first Hamiltonian form becomes
\begin{equation}
  \left(\begin{array}{c} \displaystyle
  u_t\\ \displaystyle v_t
  \end{array}
  \right) = J_0 \left(\begin{array}{c}
 \delta_u H_1 \\ \delta_v H_1
  \end{array}
  \right) = \left(
  \begin{array}{cc}
                0         & \displaystyle \frac{1}{u_{yz}} \\[4mm]
 \displaystyle - \frac{1}{u_{yz}} & \displaystyle J_0^{22}
  \end{array}
  \right)
  \left(\begin{array}{c}
   \delta_u H_1 \\ \delta_v H_1  \end{array}
  \right)
\label{sysHam1}
\end{equation}
with $J_0^{22}$ defined in (\ref{J022a}). Here $\delta_u$ and $\delta_v$ are Euler-Lagrange operators with respect to $u$ and $v$, respectively, closely related to the variational derivatives of the functional ${\cal{H}}_1 = \int\limits_{-\infty}^{+\infty}H_1\,dx dy dz$ with respect to $u$ and $v$
\cite{olv}. Here we change the notation $\mathbf{E}_\alpha$ of \cite{olv} to $\delta_{u^\alpha}$.

\section{Symmetries and integrals of motion}
\setcounter{equation}{0}
 \label{sec-symint}

Point Lie symmetries of the system (\ref{twocomp}) are determined by the following generators
\begin{eqnarray}
 & & X_1 = t\partial_t + x\partial_x + u\partial_u,\quad X_2 = \partial_t - b\,\partial_x,\quad X_3 = \partial_t
  \nonumber \\
 & & X_4 = u\partial_u + v\partial_v,\quad X_5 = f(z)\partial_u,\quad X_6 = g(y)\partial_u, \quad X_7 = h(y)\partial_y
  \label{sym}  \\
 & & X_8 = k(z)\partial_z,\; X_{cd} = \{c(t+x) + d(t-x)\}\partial_u + \{c'(t+x) + d^{\,'}(t-x)\}\partial_v \nonumber
\end{eqnarray}
where $f,g,h,k,c,d$ are arbitrary functions of a single variable and primes denote derivatives of $c$ and $d$.
The Lie subgroup of commuting symmetries is generated by the Lie subalgebra $\{X_2, X_3, X_5, X_6\}$.

We need symmetry characteristics determining symmetries in evolutionary form \cite{olv}. For the point symmetry generator of the form
$X = \xi^i\partial_{x^i} + \eta^\alpha\partial_{u^\alpha}$, where the summation over repeated indices is used, the symmetry characteristics are defined as
$\varphi^\alpha = \eta^\alpha - u^\alpha_i\xi^i$ with the subscripts $i$ denoting derivatives with respect to $x^i$. In our problem, $\alpha = 1,2$,
$u^1=u, u^2=v$, $\eta^1=\eta^u$, $\eta^2=\eta^v$, $\sum_i = \sum_{i=1}^4$, $x^1 = t, x^2 = x, x^3 = y, x^4 = z$ and $\varphi^1 = \varphi$ while $\varphi^2 = \psi$, where $\varphi$ and $\psi$ determine the transformation of $u$ and $v$, respectively.
We also use $u_t = v$ and $v_t =q$ where $q$ is the right-hand side of the second of our equations (\ref{twocomp})
\begin{equation}
q =  \frac{1}{u_{yz}}\left[u_{xx}u_{yz} - u_{xy}u_{xz} + v_yv_z + b(v_yu_{xz} - v_zu_{xy})\right].
\label{q}
\end{equation}
The symmetry characteristics become
\begin{equation}
 \varphi = \eta^u - v\xi^t - u_x\xi^x - u_y\xi^y - u_z\xi^z, \quad \psi = \eta^v - q\xi^t - v_x\xi^x - v_y\xi^y - v_z\xi^z .
\label{char}
\end{equation}
Applying the formula (\ref{char}) to the generators in (\ref{sym}), we obtain characteristics of these symmetries
\begin{eqnarray}
  & & \varphi_1 = u - tv - xu_x,\;\psi_1 = - t q - xv_x;  \nonumber \\
  & & \varphi_2 = - v + b u_x,\; \psi_2 = - q + b v_x; \quad \varphi_3 = - v,\;\psi_3 = - q  \nonumber \\
  & & \varphi_4 = u,\; \psi_4 = v;\quad \varphi_5 = f(z),\; \psi_5 = 0;\quad \varphi_6 = g(y),\;\psi_6 = 0 \nonumber\\
  & & \varphi_7 = - h(y)u_y,\;\psi_7 = - h(y)v_y;\quad \varphi_8 = - k(z)u_z,\;\psi_8 = - k(z)v_z;\nonumber\\
  & & \varphi_{cd} = c(t+x) + d(t-x),\quad \psi_{cd} = c'(t+x) + d^{\,'}(t-x).
  \label{chartab}
\end{eqnarray}
From (\ref{chartab}), it is clear that that the GH system itself is the symmetry generated by $- X_3$.

First Hamiltonian structure provides a link between symmetries in evolutionary form and integrals of motion conserved by the Hamiltonian flow (\ref{twocomp}) or in the explicitly Hamiltonian form (\ref{sysHam1})
\begin{equation}
  \left(\begin{array}{c} \displaystyle
  u_t\\ \displaystyle v_t
  \end{array}
  \right) = J_0 \left(\begin{array}{c}
  \delta_u H_1 \\ \delta_v H_1  \end{array}
  \right)
\label{Ham1form}
\end{equation}
Replacing time $t$ by the group parameter $\tau$ in (\ref{Ham1form}) and using $u_\tau = \varphi,\; v_\tau = \psi$ for symmetries in the evolutionary form,
we arrive at the Hamiltonian form of the Noether theorem for any conserved density $H$ of an integral of motion
\begin{equation}
  \left(\begin{array}{c} \displaystyle \varphi
  \\ \displaystyle \psi
  \end{array}
  \right) = J_0 \left(\begin{array}{c}
  \delta_u H \\ \delta_v H  \end{array}
  \right).
\label{Noether}
\end{equation}
To determine the integral $H$ that corresponds to a known symmetry with the characteristic $(\varphi, \psi)$ we use the inverse Noether theorem
\begin{equation}
 \left(\begin{array}{c}
  \delta_u H \\ \delta_v H  \end{array}
  \right) = K
  \left(\begin{array}{c} \varphi
  \\ \psi
  \end{array}
  \right)
\label{InvNoeth}
\end{equation}
where operator $K = J_0^{-1}$ is defined in (\ref{Kmatr}). Here (\ref{InvNoeth}) is obtained by applying the operator $K$ to both sides of (\ref{Noether}).

Let us now apply the formula (\ref{InvNoeth}) to determine integrals $H^i$ corresponding to all variational symmetries with characteristics $(\varphi_i, \psi_i)$
from the list (\ref{chartab}). Using the expression (\ref{Kmatr}) for $K$, we rewrite the formula (\ref{InvNoeth}) in an explicit form
\begin{equation}
  \left(\begin{array}{c}
  \delta_u H^i \\ \delta_v H^i  \end{array}
  \right) = \left(\begin{array}{cc}
  K_{11} & - u_{yz} \\
  u_{yz} & 0
  \end{array}
  \right) \left(\begin{array}{c} \varphi_i
  \\ \psi_i
  \end{array}
  \right)
\label{explicit}
\end{equation}
which provides the formulas for determining integrals $H^i$ for the known symmetries $(\varphi_i, \psi_i)$
\begin{equation}
 \delta_u H^i = K_{11}\varphi_i - u_{yz}\psi_i,\quad \delta_v H^i = u_{yz}\varphi_i
\label{sym_H}
\end{equation}
where $K_{11}$ is determined from (\ref{Kmatr}) to be
\begin{equation}
 K_{11} = \left(v_z + b u_{xz}\right)D_y + \left(v_y - b u_{xy}\right)D_z + v_{yz} .
\label{K_11}
\end{equation}
We always start with solving the second equation in (\ref{sym_H}) in which, due to the fact that $\varphi_i$ never contain derivatives of $v$,
$\delta_v H^i$ is reduced to the partial derivative $H^i_v$ with respect to $v$, so that the equation $H^i_v = u_{yz}\varphi_i$ is easily integrated
with respect to $v$ with the "constant of integration" $h^i[u]$ depending only on $u$ and its derivatives. Then the operator $\delta_u$ is applied to the resulting $H^i$, which involves the unknown $\delta_u h^i[u]$, and the result is equated to $\delta_u H^i$ following from the first equation in (\ref{sym_H}) to
determine $\delta_u h^i[u]$. Finally, we reconstruct $h^i[u]$ and hence $H^i$. If we encounter a contradiction, then this particular symmetry is not a variational one and does not lead to an integral.

This solution algorithm for the symmetries listed in (\ref{sym}) with characteristics (\ref{chartab}) yields the following results. $X_1$ and $X_4$ from (\ref{sym}) generate non-variational symmetries. For all other symmetries the corresponding integrals read
\begin{eqnarray}
  & & H^2 = \left\{b vu_x - \frac{1}{2} (u_x^2 + v^2) \right\}u_{yz},\nonumber \\
  & & H^3 = - \frac{1}{2} (v^2u_{yz} + u_yu_zu_{xx}) = - H_1, \quad\textrm{defined in (\ref{H1})} \nonumber \\
  & & H^5 = f(z)vu_{yz} + \frac{b}{2} f'(z)u_xu_y, \nonumber\\
  & & H^6 = g(y)vu_{yz} - \frac{b}{2} g'(y)u_xu_z  \nonumber \\
  & & H^7 = - \frac{1}{4} h(y)\left\{4 vu_yu_{yz} + b (2u_xu_yu_{yz} - u_y^2u_{xz}) \right\} \nonumber\\
  & & H^8 = - \frac{1}{4} k(z)\left\{4 vu_zu_{yz} - b (2u_xu_zu_{yz} - u_z^2u_{xy}) \right\}
   \label{int} \\
  & & H^{cd} = (c+d)vu_{yz} + \frac{(c'+d^{\,'})}{2} u_yu_z ,\quad c=c(t+x),\;d = d(t-x).  \nonumber
\end{eqnarray}

We have checked that the time derivatives of all the Hamiltonian densities $H^i$ along the flow (\ref{twocomp}) are total three-dimensional divergences which present an independent check that the corresponding functionals ${\cal{H}}^i = \int\limits_{-\infty}^{+\infty}H^i\,dx dy dz$ are indeed integrals of motion of the flow (\ref{twocomp}) subject to suitable boundary conditions. This agrees with the fact that the symmetry generators $X_2, X_5, X_6, X_7$ and $X_8$ commute with $X_3$. On the contrary, $H^{cd}$ is not conserved by the GH flow, since $X_3$ and $X_{cd}$ do not commute, but it is conserved by its own flow generated by $X_{cd}$ and also by the flows $ X_5, X_6, X_7$ and $X_8$ commuting with $X_{cd}$.
Similarly, each $H^i$ is the density of the integral of motion along all the symmetry flows with characteristics (\ref{chartab}) that commute with the corresponding $X_i$.

We also note that by replacing $v$ by $u_t$ in the expressions (\ref{int}) for $H^i$ we obtain conserved densities for the original one-component form of the general heavenly equation (\ref{gener}).

\section{Recursion operator for the two-component form of the general heavenly equation}
\setcounter{equation}{0}
 \label{sec-recurs_twocomp}

Recursion operators in a two-component form are required for constructing new Hamiltonian operators for the two-component system (\ref{twocomp}).

Recursion relations (\ref{ro1}) become
\begin{eqnarray}
& & u_{yz}(\varphi_t+\varphi_x) - (v_y+u_{xy})\varphi_z = \gamma\{u_{yz}(\psi_t+\psi_x) - (v_z+u_{xz})\psi_y\} \nonumber \\
& & u_{yz}(\varphi_t - \varphi_x) - (v_y - u_{xy})\varphi_z  = - \beta\{u_{yz}(\psi_t-\psi_x) - (v_z-u_{xz})\psi_y\}. \nonumber \\
& & \label{recurtx}
\end{eqnarray}
Now we change the notation: $\psi$ is now reserved for the second component of the symmetry characteristic, so that Lie equations read
\begin{equation}
 \left(\begin{array}{c}
 u \\
 v
 \end{array}\right)_\tau =
 \left(\begin{array}{c}
  \varphi \\ \psi
 \end{array}\right)
\label{sym2comp}
\end{equation}
where $v=u_t$ implies $\psi=\varphi_t$,
and the symmetry transformed by recursion operator $R$ is denoted by tilde:
\begin{equation}
\left(\begin{array}{c}
  \tilde\varphi \\ \tilde\psi
 \end{array}\right) = R\left(\begin{array}{c}
  \varphi \\ \psi
 \end{array}\right)
 \label{R}
 \end{equation}
where $\varphi_t = \psi$ and $\tilde{\varphi}_t = \tilde{\psi}$. Recursion relations (\ref{recurtx}) take the form
\begin{eqnarray*}
 u_{yz}(\psi+\varphi_x) - (v_y+u_{xy})\varphi_z = \gamma\{u_{yz}(\tilde{\psi}+\tilde{\varphi}_x) - (v_z+u_{xz})\tilde{\varphi}_y\}\\
 (v_y - u_{xy})\varphi_z - u_{yz}(\psi - \varphi_x) = \beta\{u_{yz}(\tilde{\psi}-\tilde{\varphi}_x) - (v_z-u_{xz})\tilde{\varphi}_y\}
\end{eqnarray*}
which in a two-component matrix form become
\begin{eqnarray}
 && \left(\begin{array}{cc}
  u_{yz}D_x-(v_y+u_{xy})D_z & u_{yz} \\
  (v_y-u_{xy})D_z+u_{yz}D_x & - u_{yz}
 \end{array}
 \right)\left(\begin{array}{c}
  \varphi \\ \psi
 \end{array}\right)\nonumber \\
 && = \left(\begin{array}{cc}
 \gamma \{u_{yz}D_x-(v_z+u_{xz})D_y\} & \gamma u_{yz} \\
 \beta \{-u_{yz}D_x+(u_{xz}-v_z)D_y\} & \beta u_{yz}
 \end{array}
 \right)\left(\begin{array}{c}
  \tilde{\varphi} \\ \tilde{\psi}
 \end{array}\right)
\label{matr_recurseq}
\end{eqnarray}
To obtain the recursion operator $R$ explicitly, we need to invert the matrix differential operator on the right-hand side of the relation (\ref{matr_recurseq}).
Note that the inverse $\left(\begin{array}{cc} e&f\\ g&h \end{array}\right)$ for the matrix $\left(\begin{array}{cc} a&b\\ c&d \end{array}\right)$
with non-commuting entries $a,b,c,d$, defined by the equation $\left(\begin{array}{cc} e&f\\ g&h \end{array}\right)\left(\begin{array}{cc} a&b\\ c&d \end{array}\right) = \left(\begin{array}{cc} 1&0\\ 0&1 \end{array}\right)$, is determined by the formulas
\begin{equation}
 e=(a-bd^{-1}c)^{-1},\; f=(c-db^{-1}a)^{-1},\; g=(b-ac^{-1}d)^{-1},\; h=(d-ca^{-1}b)^{-1}.
\label{inverse}
\end{equation}
The result for $R$ is convenient to express in terms of the two differential operators
\begin{equation}
    w = u_{yz}D_x - u_{xz}D_y,\qquad \zeta = u_{yz}D_x - u_{xy}D_z.
 \label{wzeta}
\end{equation}
Using the corresponding entries of the matrix on the right-hand side of (\ref{matr_recurseq}) for  $a,b,c,d$ in the equations (\ref{inverse}), we obtain $e,f,g,h$
\begin{eqnarray*}
   & & e = \frac{1}{2\gamma}w^{-1},\quad f = - \frac{1}{2\beta}w^{-1},\quad g = \frac{1}{2\gamma u_{yz}}(1 + v_zD_yw^{-1})
   \\    & &  h = \frac{1}{2\beta u_{yz}}(1 - v_zD_yw^{-1}).
\end{eqnarray*}
The recursion operator is explicitly defined by the relation
\begin{eqnarray}
 && R = \frac{1}{\alpha}\left(\begin{array}{cc}
   \beta   &  - \gamma \\
\displaystyle   \frac{\beta}{u_{yz}}(w + v_zD_y) &\displaystyle \frac{\gamma}{u_{yz}}(w - v_zD_y)
                        \end{array}
 \right) w^{-1} \nonumber \\
&& \times \left(\begin{array}{cc}
     \zeta - v_yD_z & u_{yz} \\
     \zeta + v_yD_z & - u_{yz}
              \end{array}
 \right)
 \label{2compR}
\end{eqnarray}
which we have multiplied by the overall constant factor $(2\beta\gamma)/\alpha$.
After performing multiplication in (\ref{2compR}) we obtain the matrix elements of $R$ explicitly
\begin{eqnarray}
   && R_{11} = w^{-1}(b \zeta + v_yD_z),\qquad R_{12} = - w^{-1}u_{yz} \nonumber \\
   && R_{21} = \frac{1}{u_{yz}} \left(v_zD_yw^{-1}v_yD_z - \zeta\right) + \frac{b}{u_{yz}}\left(v_zD_yw^{-1}\zeta - v_yD_z\right)\nonumber\\
   &&  R_{22} = b - \frac{v_z}{u_{yz}}D_yw^{-1}u_{yz},\qquad b = (\beta - \gamma)/\alpha .
 \label{matrR}
\end{eqnarray}
We do not consider the second and third recursion operators $R^2$ and $R^3$ defined in (\ref{ro2}) and (\ref{ro3}) in the two-component form because they refer to two different two-component systems.

The definition of the recursion operator in the two-component form (\ref{matrR}) contains arbitrariness related to the definition of the inverse operator $w^{-1}$ which involves addition of an arbitrary element of the kernel of $w$ defined by $w(f)=(u_{yz}D_x - u_{xz}D_y)f = 0$.
The general solution for $w(f)=0$ reads $\textrm{ker}\,w = \{f(z,u_z\}$ with an arbitrary smooth $f$. Thus, $w^{-1}$ is defined up to
the addition of an arbitrary function $f(z,u_z)$ which plays the role of an arbitrary integration constant. This arbitrariness is eliminated by the condition that not only relation $w w^{-1} = I$ is satisfied, but also $w^{-1} w = I$ has to be satisfied. Similar problems with the inversion of recursion operators for (1+1)-dimensional (systems of) PDEs were analyzed much earlier by different researchers (see \cite{G,M,S,SW} and references therein).

Let us analyze the definition of the inverse operator $w^{-1}$ in more detail. It determines a solution $f = w^{-1}g$ to the first-order ODE: $w f = u_{zy}f_x - u_{zx}f_y = g$ for any given $g$. Introduce parameter $s$ along the curves tangent to the vector field $w$, $x=x(s), y=y(s)$ on the plane $z=const$
\[w f = \frac{d f(x(s),y(s),z)}{d s} = \frac{d x}{d s} f_x + \frac{d y}{d s} f_y = u_{zy}f_x - u_{zx}f_y = g\]
which implies $\displaystyle\frac{d x}{d s} = u_{zy}$, $\displaystyle\frac{d y}{d s} = - u_{zx}$ and so
\begin{equation}
    d s = \frac{d x}{u_{zy}} = - \frac{d y}{u_{zx}}.
 \label{character}
\end{equation}
The second of these equations is the characteristic ODE with the integrals
\begin{equation}
  z = const,\quad u_z(x,y,z) = const
 \label{ints}
\end{equation}
while the first one determines the parameter $s$ along the characteristics
\begin{equation}
 s = \int\frac{d x}{u_{zy}} = - \int\frac{d y}{u_{zx}} = \frac{1}{2}\int\left(\frac{d x}{u_{zy}} - \frac{d y}{u_{zx}}\right)
 \label{s}
\end{equation}
where the integrals are taken under conditions (\ref{ints}) and the equations (\ref{character}) have been used. Under the constraints (\ref{ints}),
we have $w f = \displaystyle\frac{d f}{d s} \equiv D_s f = g$, so that $f = w^{-1}g = D_s^{-1}g = \int d s\, g$.

We define $w^{-1}$ as the integral operator $\int ds$ with $ds$ defined by (\ref{character})
\begin{eqnarray}
  w^{-1} = \left\{\int_a^x\frac{d \xi}{u_{zy}(\xi,y(\xi,c,z),z)}\right\}_{c=u_z(x,y,z)}\nonumber \\ \textrm{or}\qquad
  w^{-1} = - \left\{\int_b^y\frac{d \eta}{u_{zx}(x(\eta,c,z),\eta,z)}\right\}_{c=u_z(x,y,z)}
  \label{1/w}
 \label{w_inv}
\end{eqnarray}
where all the integrals are taken at the conditions (\ref{ints}) so that $y=y(x,c,z)$ and $x=x(y,c,z)$ are determined by the equation $u_z(x,y,z) = c = const$ and $z$ is a parameter of the integrations.
Here $a$ and $b$ are arbitrarily fixed points such that the functions $f(x,y,z)$ are subject to the boundary condition $f|_{x=a} = 0$ or $f|_{y=b} = 0$, respectively. Then for these classes of functions one can check that $w^{-1}wf(x,y,z) = f$ and also that $ww^{-1}f(x,y,z) = f$ so that $w^{-1}w = I$ and
$ww^{-1} = I$.

Indeed, using characteristic equations (\ref{character}) and definition (\ref{wzeta}) of $w$ for the first integral in (\ref{w_inv}) we have
\begin{eqnarray*}
 &\hspace*{-2pt} &w^{-1}w f(x,y,z) = \left\{\int_a^x\frac{d\xi}{u_{zy}} u_{zy}\frac{\partial}{\partial\xi}f(\xi,y,z) - \int_a^x\frac{d\xi}{u_{zy}}u_{zx}\frac{\partial}{\partial y} f(\xi,y,z)\right\}_{c=u_z} \\
  &\hspace*{-2pt} & = \underset{\{y=y(\xi,\,c,z)\}}{\int_a^x} d \xi \left\{\partial_\xi f(\xi,y,z) + \partial_\xi y\, \partial_y f(\xi,y,z)\right\}_{c=u_z} \\[1mm]
 &\hspace*{-2pt} & = \left\{\int_a^x d\xi D_\xi f(\xi,y(\xi,c,z),z)\right\}_{c=u_z} = [f(x,y(x,c,z),z)|_{c=u_z}]|_a^x \\
 &\hspace*{-2pt} & = f(x,y,z) - f(a,y,z) = f(x,y,z)
\end{eqnarray*}
for all $f$ satisfying $f|_{x=a} = 0$, so that $w^{-1}w = I$.

Next, define
\[F(x,u_z(x,y,z),z) = \left\{\int_a^x d \xi \frac{f(\xi,y(\xi,c,z),z)}{u_{zy}(\xi,y(\xi,c,z),z)}\right\}_{c=u_z(x,y,z)} \]
where we should use the equations of the characteristics (\ref{character}) for the integral. Then
\begin{eqnarray*}
& & w w^{-1} f(x,y,z) = (u_{zy}D_x - u_{zx}D_y)\left\{\int_a^x d \xi \frac{f(\xi,y(\xi,c,z),z)}{u_{zy}(\xi,y(\xi,c,z),z)}\right\}_{c=u_z(x,y,z)}\\
& & = (u_{zy}D_x - u_{zx}D_y)F(x,u_z(x,y,z),z) = u_{zy}F_x + u_{zy}F_{u_z}u_{zx} - u_{zx}F_{u_z}u_{zy} \\
& & = u_{zy}F_x = u_{zy}\frac{f(x,y,z)}{u_{zy}} = f(x,y,z)
\end{eqnarray*}
so that $ww^{-1} = I$. The check of the second definition of $w^{-1}$ in (\ref{w_inv}) could be made in a similar way.

With a more symmetric definition
\[w^{-1} = \frac{1}{2}\left\{\int_{-a}^x\frac{d \xi}{u_{zy}(\xi,y(\xi,c,z),z)} - \int_{x}^a\frac{d \xi}{u_{zy}(\xi,y(\xi,c,z),z)}\right\}_{c=u_z(x,y,z)}\]
$w^{-1}w = I$ implies the boundary condition $f(-a,y,z) = - f(a,y,z)$ for an arbitrarily fixed $a$.

\section{Second Hamiltonian structure}
\setcounter{equation}{0}
 \label{sec-2ndH}

A natural candidate for the second Hamiltonian operator $J_1$ is obtained by composing the recursion operator (\ref{2compR}) with the first Hamiltonian operator $J_1 = R J_0$ or
explicitly
\begin{equation}
\left(
\begin{array}{cc}
 J_1^{11} & J_1^{12} \\[2mm]
 J_1^{21} & J_1^{22}
\end{array}
\right) = \left(
\begin{array}{cc}
 R_{11} & R_{12} \\[2mm]
 R_{21} & R_{22}
\end{array}
\right) \left(
\begin{array}{cc}
 0 & \displaystyle \frac{1}{u_{yz}} \\[2mm]
 \displaystyle - \frac{1}{u_{yz}} & J_0^{22}
\end{array}
\right)
  \label{defJ1}
\end{equation}
where we have used the formula (\ref{Ham1}) and $J_0^{22}$ is defined in (\ref{J022a}). Utilizing also the expressions (\ref{matrR}),
we obtain the matrix elements of $J_1$ in the form
\begin{eqnarray}
  & & J_1^{11} = w^{-1},\quad J_1^{12} = \frac{b}{u_{yz}} - w^{-1} D_y\frac{v_z}{u_{yz}}\nonumber \\
  & & J_1^{21} = - \frac{b}{u_{yz}} + \frac{v_z}{u_{yz}}D_y w^{-1} \nonumber \\
  & & J_1^{22} = - \frac{1}{u_{yz}}\, \zeta \frac{1}{u_{yz}} + \frac{b^2}{u_{yz}}(\zeta - w)\frac{1}{u_{yz}}
  + \frac{b}{u_{yz}}\left(D_yv_z + v_zD_y\right)\frac{1}{u_{yz}} \nonumber \\
   & & \mbox{} - \frac{v_z}{u_{yz}} D_y w^{-1} D_y \frac{v_z}{u_{yz}},\qquad \textrm{where}\quad b = \frac{(\beta-\gamma)}{\alpha}
  \label{J_1}
\end{eqnarray}
which is explicitly skew-symmetric. Since it is not known that the recursion operator $R$ in (\ref{2compR}) is hereditary (Nijenhuis), there is no guarantee that the action of $R$ on the Hamiltonian operator $J_0$ yields again a Hamiltonian operator. Therefore, we have to check explicitly in section \ref{sec-jacobi} the Jacobi identities for the operator $J_1$ and the compatibility of the two Hamiltonian structures $J_0$ and $J_1$.

The Hamiltonian density corresponding to the general heavenly system (\ref{twocomp}) with respect to the second Hamiltonian operator $J_1$ is
\begin{equation}
    H_0 = \frac{1}{2(b^2-1)}\left[2u_x v + b (v^2 + u_x^2)\right]u_{yz} .
 \label{H0}
\end{equation}
Here we eliminate the case of the degenerate GHE with $\beta\cdot\gamma=0$, so that $b^2-1\neq 0$.
Thus, the general heavenly flow takes the bi-Hamiltonian form
\begin{equation}
  \left(\begin{array}{c} \displaystyle
  u_t\\ \displaystyle v_t
  \end{array}
  \right) = J_0 \left(\begin{array}{c}
 \delta_u H_1 \\ \delta_v H_1
  \end{array}
  \right) = J_1
  \left(\begin{array}{c}
   \delta_u H_0 \\ \delta_v H_0  \end{array}
  \right)
\label{biHam}
\end{equation}
provided that $J_1$ is proved to be a Hamiltonian operator and $J_0$ and $J_1$ are compatible.

\section{Jacobi identities for $J_1$ and compatibility of the two Hamiltonian structures $J_0$ and $J_1$}
 \setcounter{equation}{0}
\label{sec-jacobi}

Compatibility of two Hamiltonian structures means that the linear combination of the two Hamiltonian operators with arbitrary constant coefficients is also
a Hamiltonian operator. Since $J_0$ and $J_1$ both are obviously skew symmetric, the remaining problem is to prove the Jacobi identities for the mixture $J = J_1 + a J_0$ of the two Hamiltonian operators with an arbitrary constant mixing parameter $a$. Thus, we check simultaneously that $J_1$ is indeed a second Hamiltonian operator, at $a=0$, and that the two Hamiltonian operators $J_0$ and $J_1$ are compatible ($J_0$ and $J_1$ form  a \textit{Poisson pencil}) and the system (\ref{twocomp}) is \emph{bi-Hamiltonian}. Since we will use the technique of P.~Olver's book \cite{olv}, below we give a short summary of the notation and results from this book.

Let $A^l$ be the vector space of $l$-component differential functions that depend on independent and dependent variables of the problem and also on partial derivatives of the dependent variables up to some fixed order. A linear operator $J:A^l\rightarrow A^l$ is called
\textit{Hamiltonian} if its Poisson bracket
$\{\mathscr{P},\mathscr{Q}\}=\int\delta \mathscr{P}\cdot J\delta
\mathscr{Q}\ dx dy dz$ is \textit{skew-symmetric}
\begin{eqnarray}\label{skew}
  \{\mathscr{P},\mathscr{Q}\} &=& - \{\mathscr{Q},\mathscr{P}\},
\end{eqnarray}
and satisfies the \textit{Jacobi identity}
\begin{equation}\label{jacobi}
    \{\{\mathscr{P},\mathscr{Q}\},\mathscr{R}\} + \{\{\mathscr{R},\mathscr{P}\},\mathscr{Q}\} + \{\{\mathscr{Q},\mathscr{R}\},\mathscr{P}\} = 0
\end{equation}
for all functionals $\mathscr{P},\mathscr{Q}$ and $\mathscr{R}$, where $\delta$ is the variational derivative. However, the direct
verification of the Jacobi identity (\ref{jacobi}) is a hopelessly complicated computational task.  For this reason we
will use P. Olver's theory of the functional multi-vectors, in particular, his criterion (Theorem 7.8 in his book \cite{olv}):
\begin{theorem}
Let $\mathscr{D}$ be a skew-adjoint $l\times l$ matrix
differential operator and $\Theta=\frac{1}{2}\int(\omega^T\wedge\mathscr{D}\omega)\,dx dy dz$ the corresponding functional bi-vector.
Then $\mathscr{D}$ is Hamiltonian if and only if
 \begin{equation}
\mathbf{pr\,v}_{\mathscr{D}\omega}(\Theta) = 0
 \label{criterion}
 \end{equation}
where $\mathbf{pr\,v}_{\mathscr{D}\omega}$ is a prolonged evolutionary vector field with the characteristic $\mathscr{D}\omega$ defined by
\begin{equation}
  \mathbf{pr\,v}_{\mathscr{D}\omega} = \sum\limits_{i,J}D_J\left(\sum\limits_j\mathscr{D}_{ij}\omega^{j}\right)\frac{\partial}{\partial u_{J}^{i}}, \qquad
  J=0,x,y,z,xx,xy,xz\dots
  \label{PrVdef}
\end{equation}
\end{theorem}
where $u^i_0 = u^i$ and in our case $i,j = 1,2$, while $u^1=u$ and $u^2=v$. Here $\omega = (\omega^1,\omega^2) = (\eta,\theta)$ is a functional one-form corresponding to a "uni-vector" with the following property for the action of total derivatives $D_J(\omega^i)=\omega^i_J$. By definition of the space of functional multi-vectors, integrals of total divergences in $\Theta$ and in $\mathbf{pr\,v}_{\mathscr{D}\omega}(\Theta)$ always vanish. We also note that by definition of the prolonged evolutionary vector field $\mathbf{pr\,v}_{\mathscr{D}\omega}$, it commutes with total derivatives and annihilates uni-vectors
\[\mathbf{pr\,v}_{\mathscr{D}\omega}(\omega^i) = 0.\]

To check the Jacobi identities for the operator $J = J_1 + a J_0$ we set $\mathscr{D} = J$ where
\begin{equation}
 J = \left(
 \begin{array}{cc}
  w^{-1}    & \displaystyle \frac{(a+b)}{u_{yz}} - w^{-1}D_y\frac{v_z}{u_{yz}} \\
  \displaystyle - \frac{(a+b)}{u_{yz}} + \frac{v_z}{u_{yz}}D_yw^{-1} & J^{22}
 \end{array}
 \right),
 \label{J}
\end{equation}
\begin{eqnarray}
  & & J^{22} = - \frac{1}{u_{yz}}\,\zeta\frac{1}{u_{yz}} + \frac{b(a+b)}{u_{yz}}(\zeta - w) \frac{1}{u_{yz}}
 \label{J22}
 \\ & & \mbox{} + \frac{1}{u_{yz}}\left[(a+2b)D_yv_z + aD_zv_y - (a+b)v_{yz}\right]\frac{1}{u_{yz}} - \frac{v_z}{u_{yz}}D_y w^{-1}D_y \frac{v_z}{u_{yz}}
 \nonumber
\end{eqnarray}
where the linear differential operators $w$ and $\zeta$ are defined in (\ref{wzeta}), $a$ is mixing parameter in $J$ and $b = (\beta-\gamma)/\alpha$ is the single parameter of the flow (\ref{twocomp}).

We note that the matrix operator $J$ is a nonlocal one since it contains the integral operator $w^{-1}$, inverse to $w$, which is defined in (\ref{w_inv}). Nonlocal Hamiltonian structures are typical for bi-Hamiltonian systems arising from the
equations of the Monge-Amp\`ere type that describe (anti)-self-dual gravity (see, for example \cite{nns,nsky,sy,Ya}). However, so far the only type of non-locality displayed by these examples has been due to operators inverse to total derivatives, like $D_x^{-1}$, not containing variable coefficients in their definitions, so that they commute with total derivatives. The Hamiltonian operators $J_1$ and $J$ could be called \textit{essentially nonlocal} because they contain the inverse to the operator $w$, which is a combination of total derivatives with variable coefficients, and therefore $w^{-1}$ does not commute with total derivatives.
Nevertheless, all the calculations below are made with no use of such a commutativity and even without using the explicit definition (\ref{w_inv}) of $w^{-1}$,
only its consequence $w^{-1}\cdot w = 1$.

An extensive literature exists on the theory of nonlocal Hamiltonian operators in 1+1 dimensions, e.g. \cite{sk,fe,mo,se}. However, to the authors' knowledge,
no effective methods for checking the Jacobi identities exist in the multi-dimensional case.

P. Olver's criterion is formulated for matrix-differential operators. The Jacobi identities for nonlocal symmetries were shown to be correct if nonlocal variables are included in symmetries' characteristics which imply "ghost" terms in the commutators \cite{osw,olv2}. Since according to P. Olver's method nonlocal terms are automatically included in the characteristic of the evolutionary vector field $\mathbf{pr\,v}_{J\omega}$ below, with nonlocal $J$, and all such terms are canceled in the process of application of the criterion, we believe that the criterion works correctly in this more general case. Of course, a rigorous formulation of this method for checking the Jacobi identities for nonlocal Hamiltonian operators is still awaited and could be a very worthwhile project.

The bi-vector $\Theta$ in the theorem above has the form
\begin{eqnarray}
  & & \Theta = \frac{1}{2}\int (\eta,\theta)\wedge \left(
  \begin{array}{cc}
   J^{11} & J^{12} \\
   J^{21} & J^{22}
  \end{array} \right)
  \left( \begin{array}{c}
   \eta \\
   \theta
  \end{array}
  \right)\, dx dy dz
\label{Theta}
 \\   & &  \mbox{} = \frac{1}{2}\int \left(\eta\wedge w^{-1}\eta + 2\eta\wedge \frac{(a+b)}{u_{yz}}\theta - \eta\wedge w^{-1}\left(\frac{v_z}{u_{yz}}\right)_y\theta
            \right. \nonumber
 \\  & & \left. \mbox{} + \theta\wedge\frac{v_z}{u_{yz}}(w^{-1}\eta)_y + \theta\wedge\left[\frac{u_{xy}}{u_{yz}} \left(\frac{1}{u_{yz}}\theta\right)_z - \left(\frac{1}{u_{yz}}\theta\right)_x\right] \right. \nonumber
 \\ & & \left.\mbox{} + b(a+b)\theta\wedge\left[\frac{u_{xz}}{u_{yz}}\left(\frac{1}{u_{yz}}\theta\right)_y - \frac{u_{xy}}{u_{yz}} \left(\frac{1}{u_{yz}}\theta\right)_z \right] \right. \nonumber
 \\ & & \left.\mbox{} + (a+2b)\theta\wedge\frac{v_z}{u_{yz}}\left(\frac{1}{u_{yz}}\theta\right)_y + a\theta\wedge\frac{v_y}{u_{yz}}\left(\frac{1}{u_{yz}}\theta\right)_z \right. \nonumber
 \\ & & \left. \mbox{} - \theta\wedge\frac{v_z}{u_{yz}}\left\{w^{-1}\left[\left(\frac{v_z}{u_{yz}}\theta\right)_y\right]\right\}_y \right) \, dx dy dz \nonumber
\end{eqnarray}

According to P. Olver's criterion (\ref{criterion}) applied to the operator $J$, the condition for Jacobi identities to be satisfied reads
\begin{equation}
 \mathbf{pr\,v}_{J\omega}(\Theta) = \frac{1}{2}\int (\eta,\theta)\wedge\mathbf{pr\,v}_{J\omega}\wedge\left(
  \begin{array}{cc}
   J^{11} & J^{12} \\
   J^{21} & J^{22}
  \end{array} \right)
  \left( \begin{array}{c}
   \eta \\
   \theta
  \end{array}
  \right)\, dx dy dz = 0
\label{criterJ}
\end{equation}
where (\ref{Theta}) is used for $\Theta$ and $\mathbf{pr\,v}_{J\omega}$ is the prolonged vector field with the characteristic $J\omega$ which acts on each term in the integrand of (\ref{Theta}) as the evolutionary vector field. We will further skip the integral sign in the condition (\ref{criterJ}), keeping in mind the possibility of integration by parts while always omitting total divergences, and leaving out the characteristic $J\omega$ in the notation $\mathbf{pr\,v}$ for the evolutionary vector field in (\ref{criterJ}) for brevity. We also use the following formula for the action of evolutionary vector field on $w^{-1}$
\[\mathbf{pr\,v}(w^{-1}) = - w^{-1}\mathbf{pr\,v}(w) w^{-1}\]
obtained by differentiating $ww^{-1} = 1$.

The condition (\ref{criterJ}) becomes (the letter subscripts denote partial derivatives)
\begin{eqnarray}
 &\hspace*{-2pt}& (w^{-1}\eta)\wedge\mathbf{pr\,v}(w)\wedge(w^{-1}\eta) - 2\frac{(a+b)}{u_{yz}^2}\eta\wedge\mathbf{pr\,v}(u_{yz})\wedge\theta\nonumber\\
 &\hspace*{-2pt}& \mbox{} - 2(w^{-1}\eta)_y\wedge\mathbf{pr\,v}\left(\frac{v_z}{u_{yz}}\right)\wedge\theta
 + \frac{1}{u_{yz}^2}\theta\wedge\left[\mathbf{pr\,v}(u_{yz})\wedge\theta_x + \mathbf{pr\,v}(u_{xy})\wedge\theta_z\right]\nonumber\\
 &\hspace*{-2pt}& \mbox{} - 2\frac{u_{xy}}{u_{yz}^3}\theta\wedge\mathbf{pr\,v}(u_{yz})\wedge\theta_z
 + \frac{b(a+b)}{u_{yz}^2}\theta\wedge \left[\mathbf{pr\,v}(u_{xz})\wedge\theta_y - \mathbf{pr\,v}(u_{xy})\wedge\theta_z\right]\nonumber\\
 &\hspace*{-2pt}& \mbox{} + 2\frac{b(a+b)}{u_{yz}^3}\theta\wedge\left[u_{xy}\mathbf{pr\,v}(u_{yz})\wedge\theta_z
 - u_{xz}\mathbf{pr\,v}(u_{yz})\wedge\theta_y\right]\nonumber\\
 &\hspace*{-2pt}& \mbox{} + (a+2b)\theta\wedge\mathbf{pr\,v}\left(\frac{v_z}{u_{yz}^2}\right)\wedge\theta_y
 + a\theta\wedge\mathbf{pr\,v}\left(\frac{v_y}{u_{yz}^2}\right)\wedge\theta_z\nonumber\\
 &\hspace*{-2pt}& \mbox{}
 + 2\left\{w^{-1}\left[\left(\frac{v_z}{u_{yz}}\theta\right)_y\right]\right\}_y\wedge\mathbf{pr\,v}\left(\frac{v_z}{u_{yz}}\right)\wedge\theta
 \label{prvTh}
 \\ &\hspace*{-2pt}& \mbox{}
 + w^{-1}\left[\left(\frac{v_z}{u_{yz}}\theta\right)_y\right]\wedge\mathbf{pr\,v}\wedge\left\{w^{-1}\left[\left(\frac{v_z}{u_{yz}}\theta\right)_y\right]\right\}
 = 0 \;\textrm{(mod tot div)}\nonumber
\end{eqnarray}
where ``mod tot div'' means that the left-hand side of (\ref{prvTh}) equal to a total divergence is equivalent to zero and
we have simplified this equation by applying integration by parts. For example,
\[- \eta\wedge w^{-1}\mathbf{pr\,v}(w) w^{-1}\wedge\eta = (w^{-1}\eta)\wedge\mathbf{pr\,v}(w)\wedge(w^{-1}\eta)\].
The action of the evolutionary vector field $\mathbf{pr\,v}_{J\omega}$ involved in (\ref{prvTh}) is defined in terms of the matrix elements of the operator $J$ in (\ref{J}) and (\ref{J22}) according to the following rules (the letter indices $i$ and $j$ below can take the values $x, y, z$)
\begin{eqnarray}
 && \mathbf{pr\,v}_{J\omega}(u_{ij}) = D_iD_j\{J^{11}\eta + J^{12}\theta\} \nonumber\\
 && = D_iD_j\left\{w^{-1}\eta + \frac{(a+b)}{u_{yz}}\theta - w^{-1}D_y\frac{v_z}{u_{yz}}\theta\right\}, \nonumber\\
 && \mathbf{pr\,v}_{J\omega}(v_i) = D_i\{J^{21}\eta + J^{22}\theta\} = D_i\left\{-\frac{(a+b)}{u_{yz}}\eta + \frac{v_z}{u_{yz}}D_yw^{-1}\eta \right. \nonumber\\
 && \left. \mbox{} - \left(D_x - \frac{u_{xy}}{u_{yz}}D_z\right)\frac{1}{u_{yz}}\theta + \frac{b(a+b)}{u_{yz}}(u_{xz}D_y - u_{xy}D_z)\frac{1}{u_{yz}}\theta\right. \nonumber\\
 && \left. \mbox{} + \frac{1}{u_{yz}}\left[(a+2b)v_zD_y + av_yD_z + (a+b)v_{yz}\right]\frac{1}{u_{yz}}\theta
 - \frac{v_z}{u_{yz}}D_yw^{-1}D_y\frac{v_z}{u_{yz}}\theta \right\}, \nonumber\\
 && \mathbf{pr\,v}_{J\omega}(w) = \mathbf{pr\,v}_{J\omega}(u_{yz})D_x - \mathbf{pr\,v}_{J\omega}(u_{xz})D_y
 \label{Lieact}
\end{eqnarray}
where in the last line we have used the definition $w=u_{yz}D_x - u_{xz}D_y$ from (\ref{wzeta}). One should use the results (\ref{Lieact}) in order to obtain an explicit form of the equation (\ref{prvTh}).

We consider different groups of terms in the equation (\ref{prvTh}) which should separately either vanish or be reducible to a total divergence form. These are terms trilinear in $\eta$, bilinear in $\eta$, linear in $\eta$ and without  $\eta$. Now, the operator $J$ should be Hamiltonian for an arbitrary parameter $b=(\beta-\gamma)/\alpha$ of the two-component GH flow (\ref{twocomp}) and for an arbitrary mixing parameter $a$. Therefore, vanishing of terms should happen in each group separately for different dependence on the letter coefficients $a$ and $b$. Hence, each group with a given dependence on $\eta$ is divided into subgroups with the trilinear, bilinear or linear dependence on $a$ and $b$ and the terms without constant letter coefficients and all the terms in each subgroup also should vanish separately or be a total divergence. Further, terms in each subgroup are distinguished by the type of their non-locality, i.e. either trilinear, or bilinear, or linear in $w^{-1}$ or terms without $w^{-1}$ and each such subgroup should either vanish separately or be a total divergence if there no terms containing $D_x w^{-1}$. If there are such terms and all terms in the subgroup do not cancel (up to a total divergence), then first eliminate $D_x$ using the definition (\ref{wzeta}) of $w$ to obtain $D_x = \displaystyle\frac{1}{u_{yz}}(w + u_{xz}D_y)$, so that
\begin{equation}
D_x w^{-1} = \displaystyle\frac{1}{u_{yz}} + \frac{u_{xz}}{u_{yz}}D_yw^{-1},\quad \textrm{e.g.}\quad
(w^{-1}\eta)_x = \displaystyle\frac{\eta}{u_{yz}} + \frac{u_{xz}}{u_{yz}}(w^{-1}\eta)_y.
\label{D_x}
\end{equation}
Then the terms containing the first summand in $D_x w^{-1}$ as a factor should be moved into a different group, with the lesser power dependence on the non-locality $w^{-1}$, and we account only for the remaining second summands in the group while requiring this group to vanish separately. In analyzing each subgroup of terms we apply extensively integration by parts.

For example, consider all the terms in (\ref{prvTh}) trilinear in $\eta$
\begin{eqnarray*}
 &\hspace*{-7pt}& (w^{-1}\eta)\wedge(w^{-1}\eta)_{yz}\wedge(w^{-1}\eta)_x -  (w^{-1}\eta)\wedge(w^{-1}\eta)_{xz}\wedge(w^{-1}\eta)_y \\
 &\hspace*{-7pt}& = - (w^{-1}\eta)_y\wedge(w^{-1}\eta)_z\wedge(w^{-1}\eta)_x - (w^{-1}\eta)\wedge(w^{-1}\eta)_z\wedge(w^{-1}\eta)_{xy}\\
 &\hspace*{-7pt}& \mbox{} + (w^{-1}\eta)_x\wedge(w^{-1}\eta)_z\wedge(w^{-1}\eta)_y + (w^{-1}\eta)\wedge(w^{-1}\eta)_z\wedge(w^{-1}\eta)_{yx}\\
 &\hspace*{-7pt}& = - 2 (w^{-1}\eta)_x\wedge(w^{-1}\eta)_y\wedge(w^{-1}\eta)_z \\
 &\hspace*{-7pt}& = - \frac{2}{3}\left\{D_x\left[(w^{-1}\eta)\wedge(w^{-1}\eta)_y\wedge(w^{-1}\eta)_z\right]\right.\\
 &\hspace*{-7pt}& \left. \mbox{} + D_y\left[(w^{-1}\eta)_x\wedge(w^{-1}\eta)\wedge(w^{-1}\eta)_z\right]
 + D_z\left[(w^{-1}\eta)_x\wedge(w^{-1}\eta)_y\wedge(w^{-1}\eta)\right] \right\}
\end{eqnarray*}
which vanishes up to a total divergence.

In the group of terms bilinear in $\eta$ there are two subgroups of terms, either proportional to $(a+b)$ or without $a$ and $b$. First consider terms with $(a+b)$ skipping this overall factor
\begin{eqnarray*}
 && (w^{-1}\eta)\wedge\left[\left(\frac{\theta}{u_{yz}}\right)_{yz}\wedge(w^{-1}\eta)_x - \left(\frac{\theta}{u_{yz}}\right)_{xz}\wedge(w^{-1}\eta)_y\right]\\
 && \mbox{} - 2 \frac{\eta}{u_{yz}}\wedge(w^{-1}\eta)_{yz}\wedge\frac{\theta}{u_{yz}} - 2 \left(\frac{\eta}{u_{yz}}\right)_z\wedge(w^{-1}\eta)_y\wedge
 \frac{\theta}{u_{yz}}\\
 && = - (w^{-1}\eta)_y\wedge\left(\frac{\theta}{u_{yz}}\right)_z\wedge(w^{-1}\eta)_x + (w^{-1}\eta)_x\wedge\left(\frac{\theta}{u_{yz}}\right)_z\wedge(w^{-1}\eta)_y\\
 && + 2 \frac{\eta}{u_{yz}}\wedge(w^{-1}\eta)_y\wedge\left(\frac{\theta}{u_{yz}}\right)_z \\
 && = - 2\left[\frac{\eta}{u_{yz}} + \frac{u_{xz}}{u_{yz}}(w^{-1}\eta)_y\right]\wedge(w^{-1}\eta)_y\wedge\left(\frac{\theta}{u_{yz}}\right)_z\\
 && \mbox{} + 2 \frac{\eta}{u_{yz}}\wedge(w^{-1}\eta)_y\wedge\left(\frac{\theta}{u_{yz}}\right)_z = 0
\end{eqnarray*}
where we have used integrations by part and, at the last step, the equation (\ref{D_x}).

Terms bilinear in $\eta$ without $a$ and $b$ are
\begin{eqnarray*}
 &\hspace*{-13.6pt}& 2\frac{v_z}{u_{yz}^2}(w^{-1}\eta)_y\wedge(w^{-1}\eta)_{yz}\wedge\theta +
 (w^{-1}\eta)\wedge\left<\left\{w^{-1}\left[\left(\frac{v_z}{u_{yz}}\theta\right)_y\right]\right\}_{xz}\wedge(w^{-1}\eta)_y \right. \\
 &\hspace*{-13.6pt}& \left. \mbox{} - \left\{w^{-1}\left[\left(\frac{v_z}{u_{yz}}\theta\right)_y\right]\right\}_{yz}\wedge(w^{-1}\eta)_x \right>
 - \frac{2}{u_{yz}}(w^{-1}\eta)_y\wedge\left[\frac{v_z}{u_{yz}}(w^{-1}\eta)_y\right]_z\wedge\theta \\
 &\hspace*{-13.6pt}& \mbox{} + 2\left\{(w^{-1}\eta)_x\wedge(w^{-1}\eta)_{yz} - (w^{-1}\eta)_y\wedge(w^{-1}\eta)_{xz}\right\}
 \wedge w^{-1}\left[\left(\frac{v_z}{u_{yz}}\theta\right)_y\right].
\end{eqnarray*}
By integrating by parts the second and third terms, the above expression becomes
\begin{eqnarray*}
 && 2 (w^{-1}\eta)_x\wedge(w^{-1}\eta)_y\wedge\left\{w^{-1}\left[\left(\frac{v_z}{u_{yz}}\theta\right)_y\right]\right\}_z \\
 && \mbox{} + 2 [(w^{-1}\eta)_x\wedge(w^{-1}\eta)_{yz} + (w^{-1}\eta)_{xz}\wedge(w^{-1}\eta)_y]\wedge w^{-1}\left[\left(\frac{v_z}{u_{yz}}\theta\right)_y\right] \\ && = 2D_z\left\{(w^{-1}\eta)_x\wedge(w^{-1}\eta)_y\wedge w^{-1}\left[\left(\frac{v_z}{u_{yz}}\theta\right)_y\right] \right\}
\end{eqnarray*}
which vanishes up to a total derivative.

Now, as a further example, let us consider the subgroup of terms in (\ref{prvTh}) linear in $\eta$, that are bilinear in $a, b$. Terms containing nonlocality $w^{-1}\eta$ contain the overall factor $b(a+b)$ which we skip until the end of this calculation. They have the form
\begin{eqnarray*}
 &\hspace*{-18pt}& - \frac{2}{u_{yz}} (w^{-1}\eta)_y\wedge\left[\frac{u_{xz}}{u_{yz}}\left(\frac{\theta}{u_{yz}}\right)_y
 -  \frac{u_{xy}}{u_{yz}}\left(\frac{\theta}{u_{yz}}\right)_z\right]_z\wedge\theta \\
 &\hspace*{-18pt}& \mbox{} + \frac{1}{u_{yz}^2}\left[(w^{-1}\eta)_{xz}\wedge\theta_y - (w^{-1}\eta)_{xy}\wedge\theta_z + \frac{2}{u_{yz}}(w^{-1}\eta)_{yz}\wedge
 (u_{xy}\theta_z - u_{xz}\theta_y)\right]\wedge\theta
\end{eqnarray*}
Integrating by parts we obtain
\begin{eqnarray*}
 && \left(\frac{2}{u_{yz}}\right)_z(w^{-1}\eta)_y\wedge\left(\frac{u_{xz}}{u_{yz}^2}\theta_y - \frac{u_{xy}}{u_{yz}^2}\theta_z\right)\wedge\theta \\
 && \mbox{} + 2(w^{-1}\eta)_y\wedge\left[\frac{u_{xz}}{u_{yz}^2}\left(\frac{\theta}{u_{yz}}\right)_y - \frac{u_{xy}}{u_{yz}^2}\left(\frac{1}{u_{yz}}\right)_z\theta \right]\wedge\theta_z \\
 && \mbox{} + \frac{2}{u_{yz}}(w^{-1}\eta)_x\wedge\left[\left(\frac{1}{u_{yz}}\right)_y\theta_z - \left(\frac{1}{u_{yz}}\right)_z\theta_y\right]
 \wedge\theta - \frac{2}{u_{yz}^2}(w^{-1}\eta)_x\wedge\theta_y\wedge\theta_z
\end{eqnarray*}
Next we eliminate here $(w^{-1}\eta)_x$, using the formula (\ref{D_x}), and insert the coefficient $b(a+b)$, which was skipped, to obtain
\begin{equation}
 2\frac{b(a+b)}{u_{yz}^2}\left\{\eta\wedge\left[\left(\frac{1}{u_{yz}}\right)_y\theta_z\wedge\theta - \left(\frac{1}{u_{yz}}\right)_z\theta_y\wedge\theta - \frac{1}{u_{yz}}\theta_y\wedge\theta_z\right]\right\}
 \label{etagroup}
\end{equation}
so that all the non-local terms are canceled and the resulting local terms (\ref{etagroup}) should be joined with the following terms in this group which were local from the start
\begin{eqnarray*}
 && \mbox{} - 2\frac{(a+b)^2}{u_{yz}^2}\eta\wedge\left(\frac{\theta}{u_{yz}}\right)_{yz}\wedge\theta - (a+2b)(a+b)\frac{\theta}{u_{yz}^2}\wedge
 \left(\frac{\eta}{u_{yz}}\right)_z\wedge\theta_y \\
 && \mbox{} - \frac{a(a+b)}{u_{yz}^2}\theta\wedge\left(\frac{\eta}{u_{yz}}\right)_y\wedge\theta_z
\end{eqnarray*}
After integrations by part and arising cancelations the last group of terms becomes
\begin{eqnarray*}
 && 2\frac{(a+b)^2}{u_{yz}^2}\eta\wedge  \left[-\left(\frac{\theta}{u_{yz}}\right)_{yz} + \frac{\theta_{yz}}{u_{yz}}
 \right]\wedge\theta + 2b(a+b)\frac{\eta}{u_{yz}^3}\wedge\theta_y\wedge\theta_z \\
 && \mbox{} + 2(a+b)\frac{(a+2b)}{u_{yz}^2}\left(\frac{1}{u_{yz}}\right)_z\eta\wedge\theta_y\wedge\theta + 2(a+b)\frac{a}{u_{yz}^2} \left(\frac{1}{u_{yz}}\right)_y\eta\wedge\theta_z\wedge\theta
 \end{eqnarray*}
Expanding here $\left(\displaystyle\frac{\theta}{u_{yz}}\right)_{yz}$ and joining the resulting terms together with the terms in (\ref{etagroup}) we check that all the terms cancel, so that all the terms linear in $\eta$ and bilinear in $a ,b$ vanish.

In a similar way we analyze all other groups of terms in the equation (\ref{prvTh}) and make sure that each of them either vanishes or is a total divergence. The calculations are straightforward but too lengthy to be presented here.

Thus, we have proved that the Jacobi identities are satisfied for the operator $J=J_1 + a J_0$ for any constant $a$ and any coefficient $b$ of the two-component flow (\ref{twocomp}) of the general heavenly equation and since $J$ is obviously skew-symmetric, $J$ is a Hamiltonian operator. This implies that the second operator $J_1$ is indeed Hamiltonian one and that both Hamiltonian operators $J_0$ and $J_1$ are compatible and hence the equation (\ref{twocomp}) is a bi-Hamiltonian system.

\section{Higher flows}
\setcounter{equation}{0}
\label{sec-higher}

Fuchssteiner and Fokas \cite{ff} (see also \cite{sheftel} and references therein) showed that if a recursion
operator has the form $R=J_1J_0^{-1}$, where $J_0$ and $J_1$ are compatible Hamiltonian operators, then it is hereditary
(Nijenhuis), i.e. it generates an Abelian symmetry algebra out of commuting symmetries. Moreover, the formal adjoint hereditary recursion operator acting on the vector of variational derivatives of an integral yields the vector of variational derivatives of (another) integral (though under further assumptions, see e.g. Hilfssatz 4 c) in \cite{Oe}\footnote{We are grateful to an anonymous referee for this remark}). The formal adjoint recursion operator $R^\dagger$ to our original recursion operator (\ref{matrR}) has the following matrix elements
\begin{eqnarray}
& & R^\dagger_{11} = \{b\zeta + D_zv_y\} w^{-1},\quad  R^\dagger_{21} = u_{yz} w^{-1} \nonumber \\
& & R^\dagger_{12} = -\{D_zv_y w^{-1}D_yv_z - \zeta + b (\zeta w^{-1}D_yv_z - D_zv_y)\}\frac{1}{u_{yz}} \nonumber \\
& &  R^\dagger_{22} = b - u_{yz} w^{-1}D_y\frac{v_z}{u_{yz}}
  \label{hermit}
\end{eqnarray}
It acts on the vector of variational derivatives of $ H_1$ with the result
\begin{equation}
R^\dagger \left(
\begin{array}{c}
\delta_u H_1 \\ \delta_v H_1
\end{array}
\right) = R^\dagger \left(
\begin{array}{c}
v_yv_z + vv_{yz} - wu_x \\ v u_{yz}
\end{array}
\right) = \left(
\begin{array}{c}
\delta_u H_2 \\ \delta_v H_2
\end{array}
\right). \label{delth2}
\end{equation}
We reconstruct $H_2$ from this equation to be
\begin{equation}
H_2 = \frac{b}{2}\, (v^2 + u_x^2)u_{yz} 
                                        - v u_x u_z.
\label{h2}
\end{equation}
$H_2$ is a new integral. This also can be checked straightforwardly by
computing the total time derivative of $H_2$ along the flow (\ref{twocomp}) and showing that it is a total divergence.

Since $J_1 = R J_0$, we have $R=J_1J_0^{-1}$, so that $R^\dagger = J_0^{-1}J_1$ and therefore (\ref{delth2}) can be rewritten as
\begin{equation}
  \left(\begin{array}{c} \displaystyle
  u_\tau\\ \displaystyle v_\tau
  \end{array}
  \right) = J_0 \left(\begin{array}{c}
 \delta_u H_2 \\ \delta_v H_2
  \end{array}
  \right) = J_1
  \left(\begin{array}{c}
   \delta_u H_1 \\ \delta_v H_1  \end{array}
  \right) =
  \left(\begin{array}{c} \displaystyle b v - u_x
   \\ \displaystyle b q - v_x
  \end{array}
  \right)
\label{H2flow}
\end{equation}
where $q$ is defined in (\ref{q}). This is still a local point symmetry which is a linear combination of symmetries
$(\varphi_2, \psi_2)$ and $(\varphi_3, \psi_3)$ from the list (\ref{chartab}).

A nontrivial result is obtained by applying $J_1$ to the vector of
variational derivatives of $H_2$
\begin{equation}
  \left(\begin{array}{c} \displaystyle
  u_\tau\\ \displaystyle v_\tau
  \end{array}
  \right) = J_1 \left(\begin{array}{c}
 \delta_u H_2 \\ \delta_v H_2
  \end{array}
  \right)
\label{J1H2}
\end{equation}
which yields
\begin{eqnarray}
 u_\tau  &=& (1 + b^2) v - 2 b u_x - \frac{b}{2} w^{-1}(vv_{yz} + v_yv_z)  \nonumber \\
 v_\tau  &=& (1 + b^2)\left(q + \frac{v_yv_z}{u_{yz}}\right) - 2 b v_x
 \label{J1H2flow}
\end{eqnarray}
where the right-hand side is a nonlocal symmetry characteristic.
Here the local integral $H_2$ generates a nonlocal symmetry.

We can repeat the procedure by applying $R^\dagger$ to variational derivatives of $H_2$
\begin{equation}
R^\dagger \left(
\begin{array}{c}
\delta_u H_2 \\ \delta_v H_2
\end{array}
\right) = \left(
\begin{array}{c}
\delta_u H_3 \\ \delta_v H_3
\end{array}
\right)
 \label{delth3}
\end{equation}
to determine the next Hamiltonian $H_3$, which will be used together with $J_1$ to generate further nonlocal symmetry
\begin{equation}
  \left(\begin{array}{c} \displaystyle
  u_\tau\\ \displaystyle v_\tau
  \end{array}
  \right) = J_1 \left(\begin{array}{c}
 \delta_u H_3 \\ \delta_v H_3
  \end{array}
  \right)
\label{J1H3}
\end{equation}
and so on.

In a similar way we can construct higher integrals and corresponding higher flows by applying the adjoint recursion
operator $R^\dagger$ to the variational derivatives of all the integrals constructed in section \ref{sec-symint}.

\section{Conclusion}

 We have constructed two additional Lax pairs and three nonlocal recursion operators for the general heavenly equation (GHE) obtained by the splitting of the Lax pairs with respect to the spectral parameter. Converting GHE to a two-component evolutionary form, we have discovered Lagrangian, symplectic and Hamiltonian structures of this GH system. We have determined all local Lie point symmetries of the GH system and, using the inverse Noether theorem in Hamiltonian form, we obtained Hamiltonians generating all the variational (Noether) point symmetries. These Hamiltonians are integrals of the motion along the general heavenly flow if the symmetry flows generated by them commute with the GH flow. Each Hamiltonian generating a variational point symmetry flow is conserved along each point symmetry flow that commutes with the flow generated by the Hamiltonian under study. Converting the GH system back to the general heavenly equation in the one-component form, one could obtain integrals of GHE as reductions of the integrals for the system. We have converted the first recursion operator to a matrix $2\times 2$ form appropriate for our two-component evolutionary system, while the other two operators refer to different two-component systems obtained from the first one by permutations, so that we end up with a single nonlocal recursion operator for the first GH system. Composing the recursion operator $R$ with the first Hamiltonian operator $J_0$ we have obtained the second Hamiltonian operator $J_1=RJ_0$ and found the corresponding Hamiltonian density generating the two-component flow of GHE. We have checked the Jacobi identities for $J_1$ and compatibility of the two Hamiltonian structures $J_0$ and $J_1$, so that our recursion operator is hereditary (Nijenhuis). In doing this, we have used P. Olver's theory of functional multi-vectors and showed that it works nicely even if being applied to nonlocal operators. This seems to be a well-founded conjecture because all nonlocal terms are canceled while applying the Olver's criterion, the rigorous generalization to nonlocal Hamiltonian operators being still awaited. Under this conjecture, we have shown that the GHE two-component flow is a bi-Hamiltonian system integrable in the sense of Magri.
Using the formal adjoint recursion operator $R^\dagger$, we demonstrate how to generate infinite series of Hamiltonians generating higher nonlocal symmetry flows of the GH system.

A detailed study of the hierarchies of nonlocal symmetries and conservation laws of the general heavenly system deserves more attention and is currently in progress.

\section*{Acknowledgments}

The research of M. B. Sheftel is partly supported by the research grant
from Bo\u{g}azi\c{c}i University Scientific Research Fund (BAP),
research project No. 11643. \\ We are grateful to an anonymous referee for numerous useful remarks
which contributed to the improvement of our paper.

\end{document}